\pdfoutput=1
\documentclass[twocolumn,prl,showpacs,amsmath,amssymb,superscriptaddress,floatfix]{revtex4}

\usepackage{bm}
\usepackage{graphicx}
\usepackage{amssymb}
\usepackage{amsmath}
\usepackage{xcolor}

\begin{document}

\title{Vanishing RKKY interactions in Ce-based cage compounds}

\author{A. M.  Konic}  
\affiliation{Department of Physics, Kent State University, Kent, Ohio, 44242, USA}

\author{Y. Zhu}
\affiliation{Department of Physics, University of California at San Diego, La Jolla, CA 92903, USA}
\affiliation{Center for Advanced Nanoscience, University of California, San Diego, La Jolla, California 92093, USA}

\author{A. J. Breindel} 
\affiliation{Department of Physics, University of California at San Diego, La Jolla, CA 92903, USA}
\affiliation{Center for Advanced Nanoscience, University of California, San Diego, La Jolla, California 92093, USA}

\author{Y. Deng}
\affiliation{Department of Physics, University of California at San Diego, La Jolla, CA 92903, USA}
\affiliation{Center for Advanced Nanoscience, University of California, San Diego, La Jolla, California 92093, USA}

\author{C. M. Moir}
\affiliation{Department of Physics, University of California at San Diego, La Jolla, CA 92903, USA}
\affiliation{Center for Advanced Nanoscience, University of California, San Diego, La Jolla, California 92093, USA}

\author{M. B. Maple}
\affiliation{Department of Physics, University of California at San Diego, La Jolla, CA 92903, USA}
\affiliation{Center for Advanced Nanoscience, University of California, San Diego, La Jolla, California 92093, USA}

\author{C. C. Almasan}
\affiliation{Department of Physics, Kent State University, Kent, Ohio, 44242, USA}

\author{M. Dzero}  
\affiliation{Department of Physics, Kent State University, Kent, Ohio, 44242, USA}

\date{\today}

\begin{abstract}
We report the results of thermodynamic measurements in external magnetic field of the cubic Ce-based cage compounds Ce$T_{2}$Cd$_{20}$ ($T$ = Ni,Pd). Our analysis of the heat-capacity data shows that the $\Gamma_7$ doublet is the ground state multiplet of the Ce$^{3+}$ ions. Consequently, for the $\Gamma_7$ doublet it can be theoretically shown that the Ruderman-Kittel-Kasuya-Yosida (RKKY) interaction between the localized Ce moments mediated by the conduction electrons, must vanish at temperatures much lower than the energy separating the ground state doublet from the first excited $\Gamma_8$ quartet. Our findings provide an insight as to why no long range order has been observed in these compounds down to temperatures in the millikelvin range. 

\end{abstract}

\maketitle

\section{Introduction}
The Doniach phase diagram remains one of the main conceptual frameworks used to get insight into a variety of physical phenomena in materials that contain a lattice of magnetic moments, originating from the partially filled $f$-electronic orbitals, interacting with the spins of the itinerant electrons \cite{Doniach1977}.
It describes the competition between the Kondo screening of the local moments, accounted for by an energy scale $T_K$, and the tendency for the magnetic moments to form long-range magnetic order, driven by the RKKY interaction with the energy scale $T_{\textrm{RKKY}}$. Of special interest is the transition between the two regimes defined by $T_K<T_{\textrm{RKKY}}$ and $T_K>T_{\textrm{RKKY}}$, which is thought to be characterized by a quantum critical point \cite{ColemanReview}. Whether or not a system is tuned to a quantum critical regime by pressure, magnetic field, or disorder, typically one observes the formation of a long range order (usually antiferromagnetism or superconductivity) at some temperature. Hence, it is a very rare and surprising occurrence that no long range order will be observed in metallic systems with an underlying cubic (or tetragonal) crystal structure and well-formed local magnetic moments down to temperatures in the millikelvin range. Furthermore, in this case one may conclude that in such Kondo lattice systems RKKY interactions must be either vanishingly small or even completely absent. 

In recent years, the family of the ternary compounds with chemical composition R$T_{2}X_{20}$ ($R$ = rare earth metal, $T$ = transition metal, and $X$ = Al, Zn or Cd) have introduced a new platform to explore the physical phenomena where, on one hand, electron-electron correlations play a dominant role while, on the other hand, they do not necessarily show a tendency to exhibit long range order. It is also important to emphasize that this tendency is not driven by geometric frustration since they have a cubic crystal structure. Moreover, these materials exhibit a few hallmark features such as (\emph{i}) rare earth ions are located at sites within cages of surrounding $X$ ions \cite{Niemann1995,White2015} and (\emph{ii}) fairly large separations between rare earth ions, which benefits the formation of the localized magnetic moments \cite{Ruderman1954,Kasuya1956,Yosida1957,White2015}.
For example, in the compounds Ce$T_{2}X_{20}$ ($X$ = Al, Zn), the Ce-Ce separation is found to be ~$6.2-6.4$ \AA. Lastly we note that in compounds with $X$ = Al, there is sufficiently strong hybridization between the 4$f$ and conduction electrons to prevent the local moment formation (despite the Ce-Ce distance exceeding the Hill Limit) \cite{Moze1998,Kangas2012,White2015}. On the contrary, for $X$ = Zn the hybridization is weaker and the compounds show local moment formation at high temperatures \cite{hill1970plutonium,Isikawa2013,Swatek2013,White2015}.

In this work, we present results of heat capacity measurements performed on two Ce based cage compounds, CeNi$_{2}$Cd$_{20}$ and CePd$_{2}$Cd$_{20}$, in magnetic fields ranging from 0 to 14 T and temperatures down to 0.4 K. Clear Schottky-type anomalies around 7 K, accompanied by an upturn at very low temperatures that becomes a clear peak in higher magnetic fields, are observed and analyzed to obtain information regarding the multiplet structure and splitting under the influence of the magnetic field. As it turns out, this information is crucial for obtaining the estimate for the strength of the RKKY interaction between the local moments. The value of the Sommerfeld coefficient $\gamma\approx 30$ mJ/(mole$\cdot$K$^2$) indicates a moderate degree of the hybridization between the localized $f$- and itinerant $d$-orbitals of the cerium ions. Our main finding is that, in view of the cubic symmetry of the crystalline electric fields, the ground state multiplet of the Ce$^{3+}$ ion \cite{White2015} is a Kramers doublet ($\Gamma_7$) separated from the first excited quartet ($\Gamma_8$) by $\Delta\approx 27$ K. 
As a direct consequence of this, the RKKY interaction between the neighboring Ce moments becomes identically zero at temperatures $T\ll \Delta$. In other words, both CeNi$_{2}$Cd$_{20}$ and CePd$_{2}$Cd$_{20}$ are Kondo lattice systems with vanishingly small RKKY interactions at low temperatures.  This result also means that magnetic long range order may only be driven by super-exchange interactions between the Ce ions and, therefore, must necessarily be antiferromagnetic, which is indeed the case for CeAl$_2$ \cite{MapleCeAl2}. It is unlikely to be the case for  CeT$_{2}$Cd$_{20}$ ($T$=Ni,Pd) due to the fairly large separation between the cerium ions. 

\section{Experimental Details}
Single crystals of CeNi$_2$Cd$_{20}$ and CePd$_2$Cd$_{20}$ were grown using a molten cadmium flux as described in Refs.~\cite{White2015,Burnett2014}.  Analysis of the powder x-ray diffraction (XRD) patters obtained via a Bruker D8 Discover x-ray diffractometer was done to determine the crystal structure and quality of the single crystals \cite{Yazici2015}. This analysis showed both samples are single phase crystals without any indication of impurity phases \cite{Yazici2015}. The structure for both samples was determined to be the CeCr$_{2}$Al$_{20}$-type cubic structure, having a space group of $Fd\Bar{3}m$ \cite{White2015,Burnett2014}.

We performed heat capacity measurements on a CeNi$_{2}$Cd$_{20}$ sample with a mass of 3.1 mg and a CePd$_{2}$Cd$_{20}$ sample with a mass of 6.8 mg, using the He-3 option for a Quantum Design Physical Property Measurement System (PPMS) employing a standard thermal relaxation technique.  For better thermal contact between the samples and measurement platform, the contact surface of each sample was polished with sand paper. The measurements were performed in magnetic fields $B$ ranging from 0 to 14 T applied along the $c$-axis and over a temperature range of 0.40 K $\le T \le 35$ K.

\section{Experimental Results}
\subsection{Specific heat in zero magnetic field}
In Fig.~\ref{phononsub} and its inset, we show the temperature dependence of the specific heat normalized by temperature, $C(T)/T$, in zero magnetic field for single crystal samples of CeNi$_{2}$Cd$_{20}$ and CePd$_{2}$Cd$_{20}$, respectively. In order to extract the $4f$ electron contributions to the specific heat, we first subtracted the phonon contribution in each sample. For CeNi$_{2}$Cd$_{20}$, this was done using data for the non-magnetic analog compound LaNi$_{2}$Cd$_{20}$, while for CePd$_{2}$Cd$_{20}$, we estimated the phonon contribution from the Debye model as done in Ref.~\cite{Konic2021}, using the following expression:
\begin{figure}[t]
\includegraphics[width=2.75in]{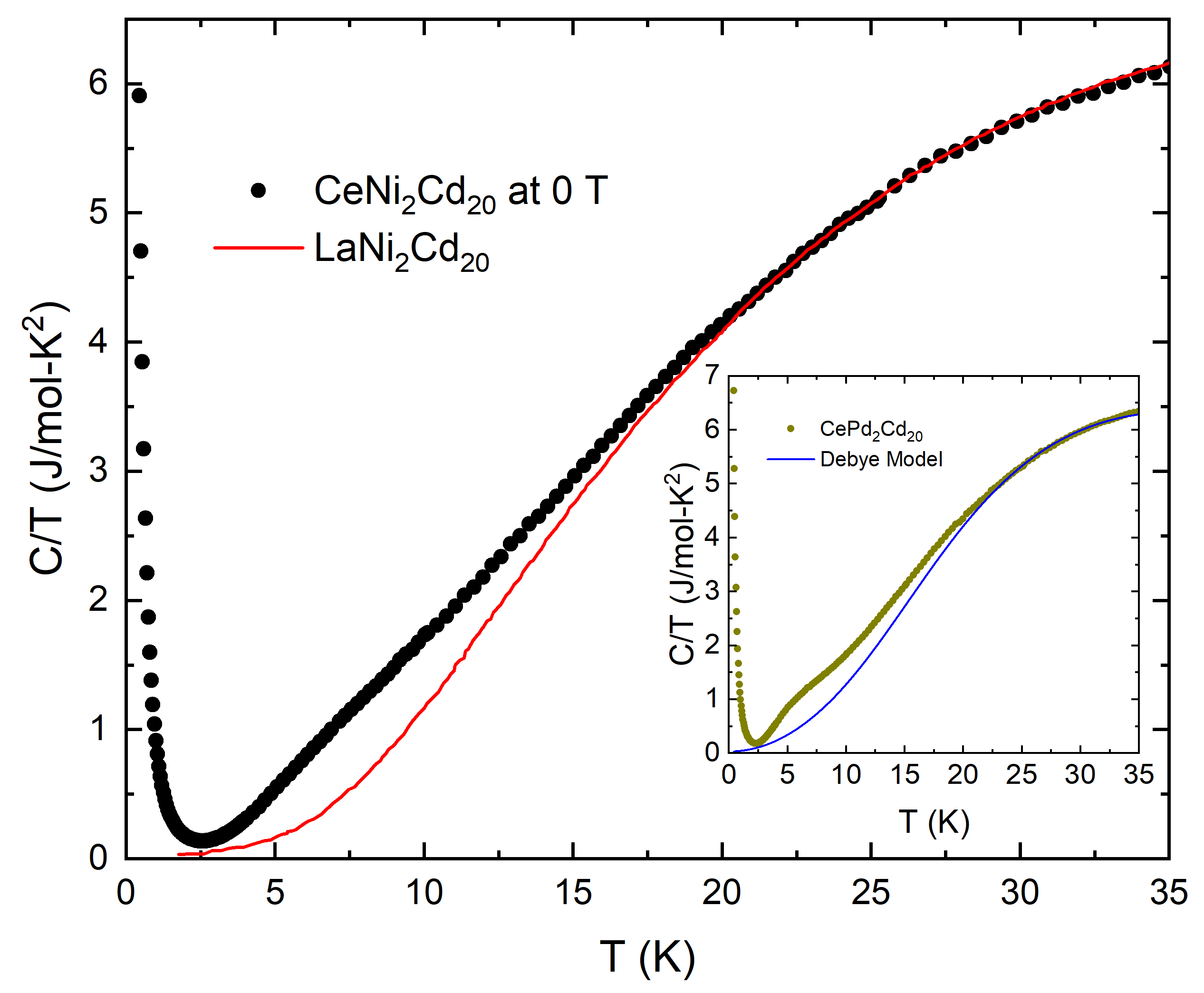}
\caption{Specific heat normalized by temperature ($C/T$) for the compounds CeNi$_{2}$Cd$_{20}$ and CePd$_{2}$Cd$_{20}$ in zero magnetic field. The non-magnetic analog LaNi$_{2}$Cd$_{20}$ was used to subtract the phonon contribution to the specifc heat for CeNi$_{2}$Cd$_{20}$, while an estimation using the Debye model was used for CePd$_{2}$Cd$_{20}$.}
\label{phononsub}
\end{figure}

\begin{equation}
    C(T)=\gamma T+{9{ N}k_B}\left(\frac{T}{\theta_D}\right)^3\int\limits_0^{{\theta_D}/{T}}\frac{x^4e^{x}dx}{(e^x-1)^2}.
    \label{phonon}
\end{equation}
Using Eq. (\ref{phonon}) to fit the data at higher temperatures, we obtained $\gamma \approx 30$ mJ/mol-K$^{2}$, and $\theta _{D} \approx 139$ K. After subtraction of the phonon contribution, we analyzed the resulting $4f$ electron contribution to the specific heat.

In the main panels of Figs. \ref{0field}(a) and 2(b), we show the resulting $4f$ electron contribution to the heat capacity after subtracting the phonon contribution for the CeNi$_{2}$Cd$_{20}$ and CePd$_{2}$Cd$_{20}$, respectively. We observe a Schottky type anomaly around $T\approx 7$ K in both compounds. To analyze our data, we recall that the Ce$^{3+}$ valence configuration \cite{White2015} corresponds to a total angular momentum $J=5/2$. In the crystalline cubic environment, the $(2J+1)$-fold degenerate multiplet splits into a $\Gamma_8$ quartet and a $\Gamma_7$ Kramers doublet \cite{LEA1962}. It is important for our subsequent discussion to emphasize that it is not clear \emph{a priori} which of the two multiplets has a lower energy. For example, we can recall a similar situation in cubic SmB$_6$ and SmS: in samarium hexaboride the ground state multiplet of Sm$^{3+}$ is $\Gamma_8$ (see Ref. \cite{Dzero2016}), while in samarium sulfide the ground state multiplet of Sm$^{3+}$ is a $\Gamma_7$ Kramers doublet \cite{Mazzone2020}. Therefore, we tried to perform the fits of the Schottky anomaloy without making explicit assumptions about which multiplet is a ground state one. Specifically, we use the following expression to analyze the heat capacity data:

\begin{figure}[t]
    \centering
    \includegraphics[width=2.75in]{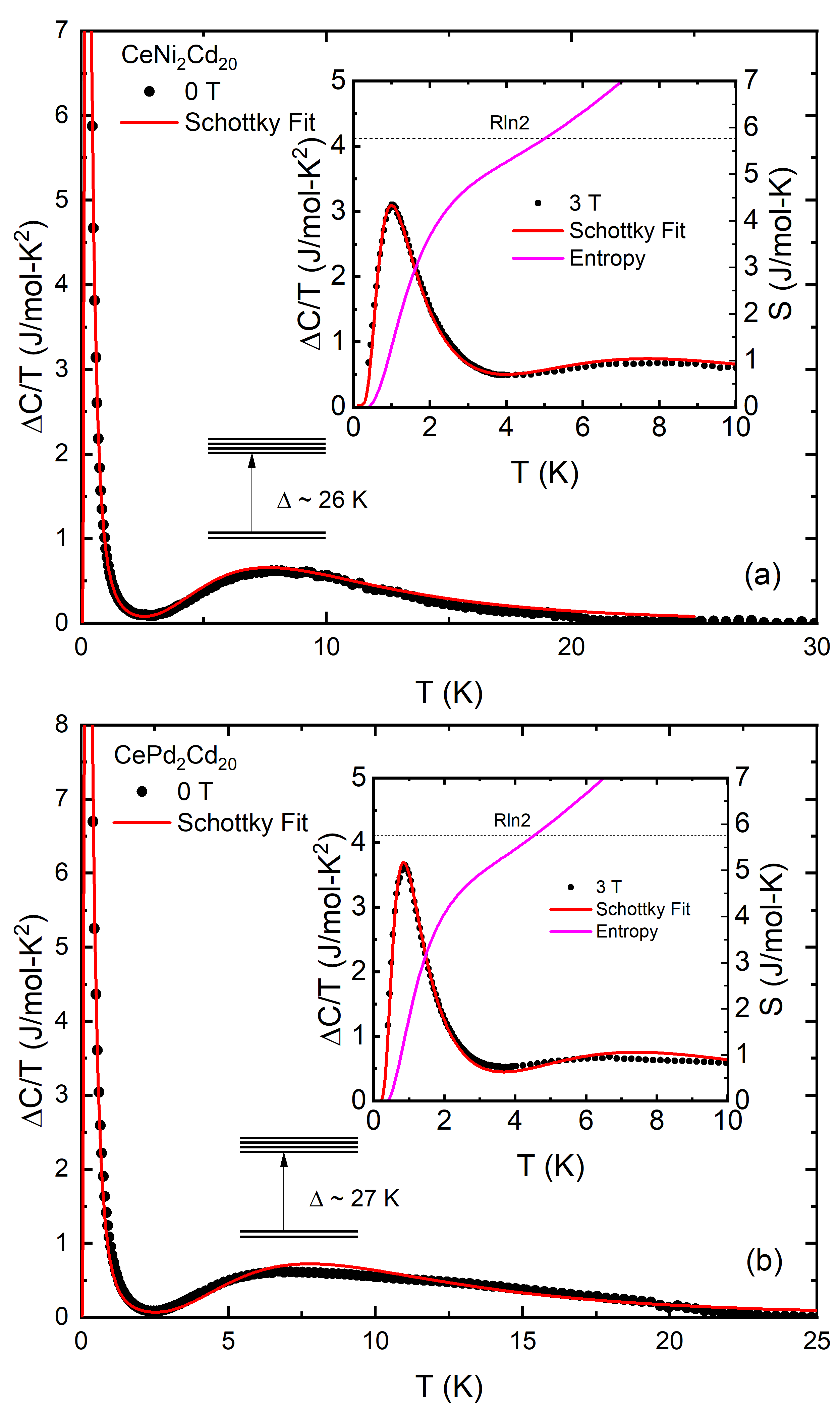}
    \caption{Electronic specific heat normalized by temperature ($\Delta C/T$) for CeNi$_{2}$Cd$_{20}$ and CePd$_{2}$4Cd$_{20}$. Both compounds exhibit a Schottky type anomaly in the heat capacity at temperatures in the vicinity of 7 K due to the crystalline electric field (CEF) splitting of the $J = 5/2$ multiplet.}
    \label{0field}
\end{figure}
\begin{equation}\label{schottky}
\frac{C_{\textrm{el}}(T)}{T} = \gamma +
\frac{A\Delta^2g_ag_be^{-\Delta/T}}{T^3\left(g_a+g_be^{-\Delta/T}\right)^2}.
\end{equation}
Here $\gamma$ is the Sommerfield coefficient, $A$ is a phenomenological parameter which describes the degree of hybridization between Ce ions and conduction band electrons \cite{Konic2021}, g$_{a}$ and g$_{b}$ are the degeneracies of the ground and first excited states, respectively, and $\Delta$ is the energy separation between them. Best fits for these compounds were obtained for values of $g_{a} = 2$, $g_{b} = 4$, $\gamma \approx 30$ mJ/mol-K$^{2}$, and $\Delta \approx 27$ K. These values correspond to energy level structure depicted in the schematic of the main panels of Figs. ~\ref{0field}, wherein the 6-fold degenerate multiplet splits into a quartet and doublet, with the doublet being the ground state configuration.

\subsection{Specific heat in external magnetic field}
In order to investigate the response of the low temperature upturn in response to an applied magnetic field, we measured the specific heat in external magnetic field up to 14 Tesla. We then performed fits using Eq. (2), with $g_{a} = g_{b} = 1$, under the assumption that this peak is due to the ground state doublet having its degeneracy lifted. The same assumption was used for the first excited quartet. The insets of Figs. \ref{0field} show these fits for 3 T and show strong agreement with the previously discussed $\Gamma_{7}$ ground state scenario.

In order to confirm this multiplet structure, we calculated the entropy for both compounds using:
\begin{equation}
    S(T) = \int\limits_0^T\frac{ C_{\textrm{el}}(T')}{T'}dT'- \gamma T,
    \label{entropy}
\end{equation}
as seen in the insets of Figs.~\ref{0field}. In both compounds, the entropy reaches a value of R$\ln2$ around 5 K, which suggests that the ground state of the six-fold degenerate multiplet is, indeed, the $\Gamma_7$ doublet.

\begin{figure}[t]
    \centering
    \includegraphics[width=3.25in]{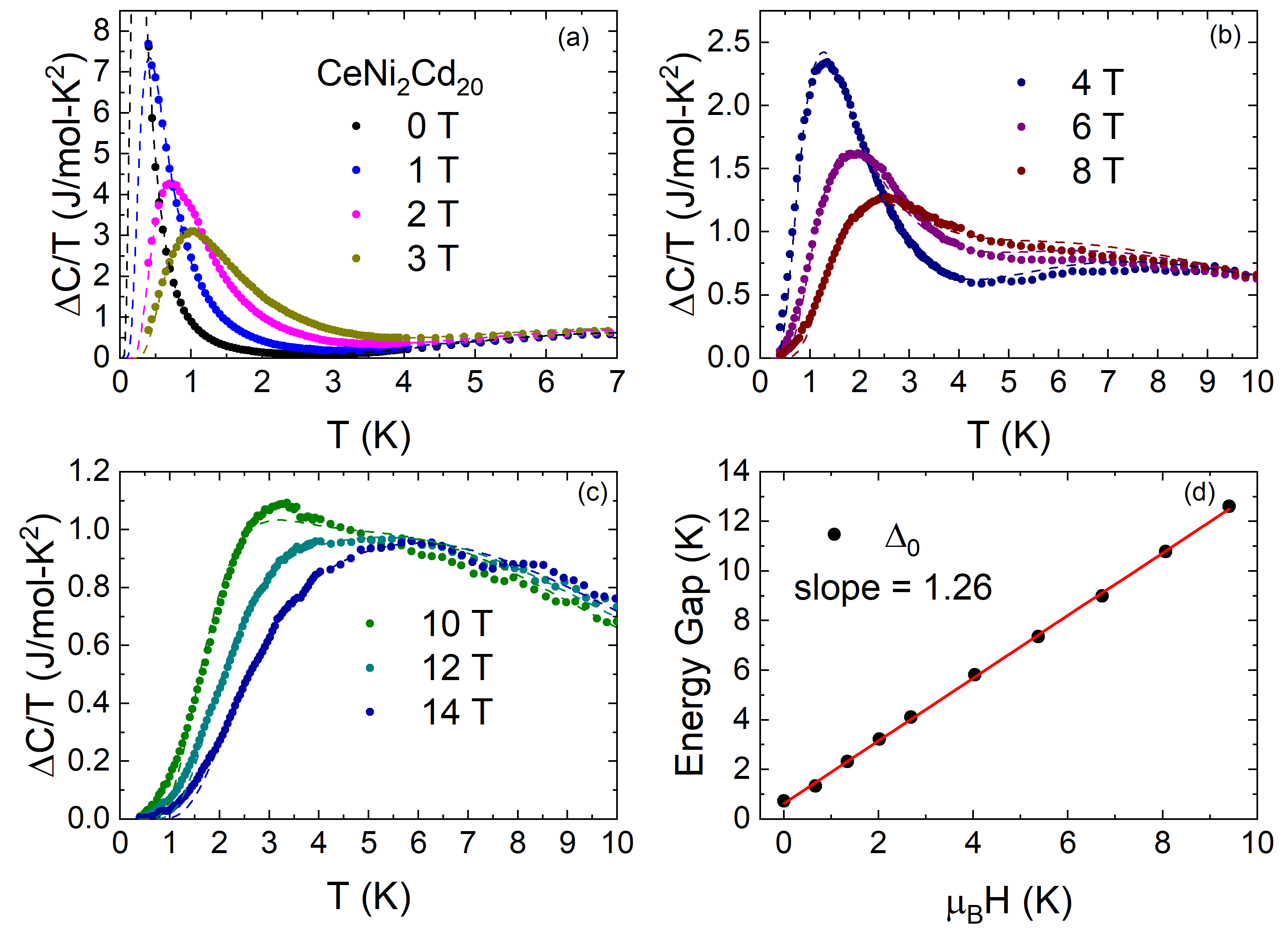}
    \caption{(a-d) Electronic specific heat normalized by temperature ($C_{\textrm{el}}/T$) for CeNi$_{2}$Cd$_{20}$ in magnetic fields ranging from 0-14 T. (d) Energy gap ($\Delta_{0}$) between the two singlets that make up the ground state doublet as a function of applied magnetic field.}
    \label{gapfitsNi}
\end{figure}

\begin{figure}
    \centering
    \includegraphics[width=3.25in]{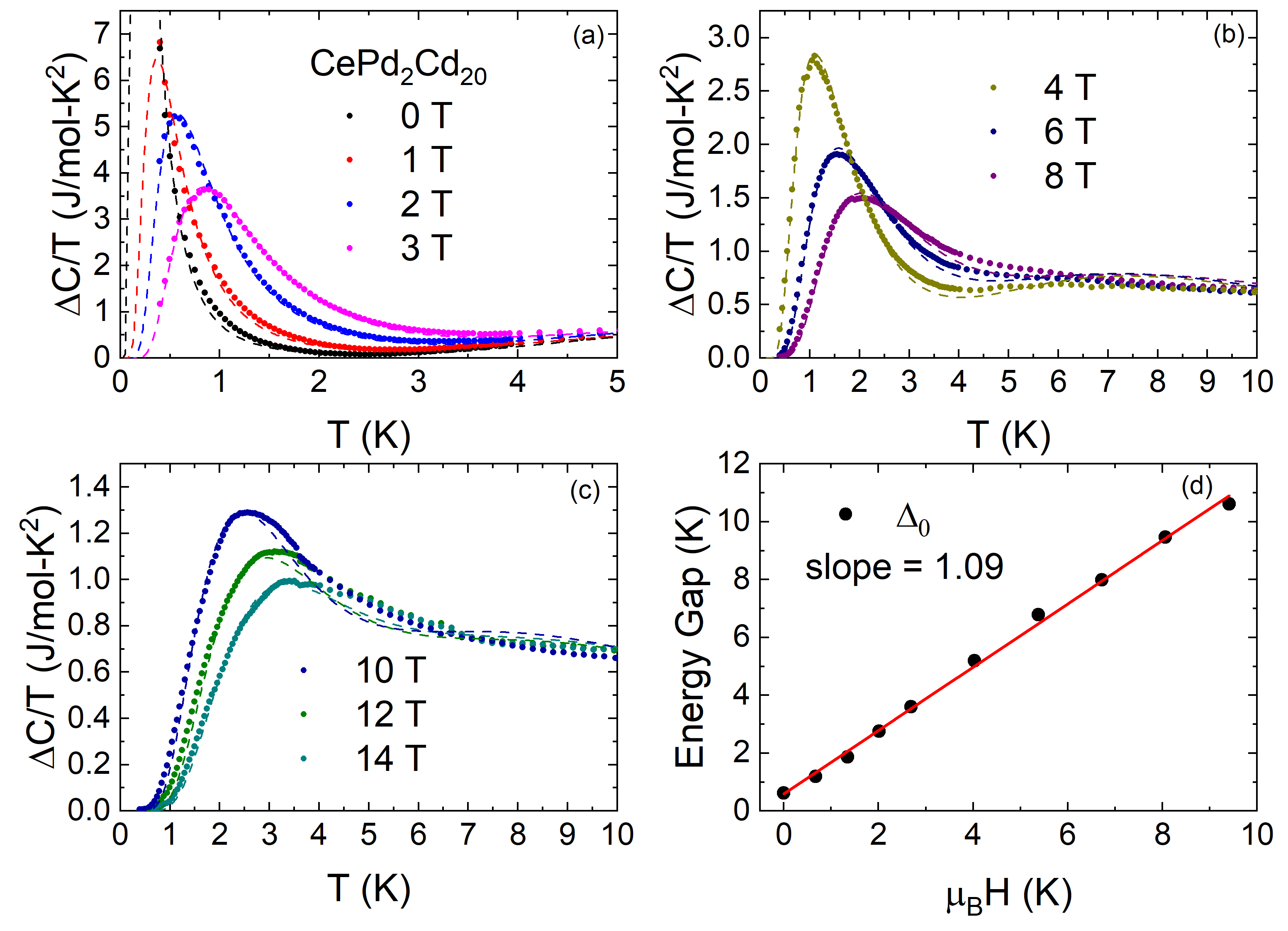}
    \caption{(a-d) Electronic specific heat normalized by temperature ($C_{\textrm{el}}/T$) for CePd$_{2}$Cd$_{20}$ in magnetic fields. (d) Energy difference ($\Delta_{0}$) between the two singlets that make up the ground state $\Gamma_7$ doublet as a function of applied magnetic field.}
    \label{gapfitsPd}
\end{figure}

Panels (a-c) of Figs. \ref{gapfitsNi} and \ref{gapfitsPd} which display the electronic specific heat for both samples revealing how a peak in $C(T)/T$ evolves under the influence of an applied magnetic field. These data also show that at in a field of $\approx 2$ T, the low temperature upturn becomes a well defined peak in the temperature range of our measurements, and then continues to shift to higher temperatures and broaden with increasing  field. We performed fits to these data, shown in Figs. \ref{gapfitsNi} and \ref{gapfitsPd}, using Eq. \eqref{schottky}, with g$_{a}$ = g$_{b}$ = 1, as we have discussed above. 

Using these fits, we then extracted the value for the energy separation between the two levels $\Delta_{0}$, to observe its behavior in magnetic field.
Panel (d) of Figs. \ref{gapfitsNi} and \ref{gapfitsPd} show plots of the energy level splitting $\Delta_0$ for the ground state $\Gamma_7$ doublet vs applied magnetic field ($H$) for both compounds. We observe that the slope $\Delta_0/\mu_BH$ in both of these plots is lower than what one would expect for CeNi$_{2}$Cd$_{20}$ and CePd$_{2}$Cd$_{20}$. Indeed, since the state vector
\begin{equation}\label{Gamma7}
\mid\Gamma_7,\pm\rangle=\sqrt{\frac{5}{6}}\mid\mp 3/2\rangle-\sqrt{\frac{1}{6}}\mid\pm 5/2\rangle,
\end{equation}
and the gyromagnetic factor is $g=6/7$, one would expect to find $\Delta_0/\mu_BH\approx 1.4$, while experimentally we find $(\Delta_0/\mu_BH)_{\textrm{exp}}\approx1.26$ and $(\Delta_0/\mu_BH)_{\textrm{exp}}\approx1.09$ for CeNi$_{2}$Cd$_{20}$ and CePd$_{2}$Cd$_{20}$, respectively. This difference is likely due to a mismatch in ionic sizes of Ni and Pd ions, so that larger ionic size of Pd produces stronger hybridization between Ce $f$ and $d$ orbitals and, as a result, lower occupation numbers of Ce $f$-levels. Spectroscopic measurements, however, are needed to verify this statement. This result signals the reduction in the effective cerium moment induced by the interaction with the conduction electrons, which we find to be consistent with an enhanced value of the Sommerfeld coefficient $\gamma\approx 30$ mJ/(mole$\cdot$K$^2$). 

\section{Discussion}
Let us first discuss some of the consequences of having $\Gamma_7$ as a ground state multiplet. Specifically, we consider the expression for the RKKY interaction between the two magnetic moments ${\mathbf S}_1$ and ${\mathbf S}_2$ separated by a distance $R_{12}$ from each other in a one-band metal (in the units of $\hbar=1$)\cite{Pedro2000}:
\begin{equation}\label{RKKY}
H_{\textrm{RKKY}}=\frac{2m(Jp_F^2)^2}{(2\pi)^3}F(2p_FR_{12})\left({\mathbf S}_1\cdot{\mathbf S}_2\right),
\end{equation}
where $F(x)=\cos(x)/x^3$, $p_F$ is the Fermi momentum, $m$ is the effective mass of a conduction electron, and $J$ is the exchange constant between the spin of the conduction electron and local moment. It is important to recall that magnetic moments ${\mathbf S}_{1,2}$ have an orbital content. In their seminal paper, Coqblin and Schrieffer \cite{Coqblin1969} have shown that for $J=5/2$ multiplets of Ce$^{3+}$ ions, the spherically symmetric expression (\ref{RKKY}) is modified due to an inherent anisotropy imposed by the cubic crystalline environment and has the following form:
\begin{equation}\label{RKKYM}
\begin{split}
H_{\textrm{RKKY}}&=\left(\frac{m_{\textrm{eff}}}{\pi^3}\right)J^2p_F^4\frac{\cos(2p_Fr)}{(2p_Fr)^3}\\&\times\sum\limits_{MM'}\phi(M)\phi(M')
\mid 1M,2M'\rangle\langle 1M',2M\mid.
\end{split}
\end{equation}
Here the summation is performed over the states of the Ce$^{3+}$ multiplet and $\phi(M)$ are the corresponding form factors:
\begin{equation}\label{FormFactors}
\phi(\pm 1/2)=3, \quad \phi(\pm3/2)=\phi(\pm5/2)=0.
\end{equation}
From Eqs. (\ref{RKKYM}, \ref{FormFactors}) above we can immediately check that for the case of the $\Gamma_7$ doublet, Eq. (\ref{Gamma7}), the RKKY interaction is identically zero . The immediate consequence of this result is the absence of long-range magnetic order, which is typically mediated by the RKKY interaction in Kondo lattice systems. 

There, of course, exists the possibility of forming long-range antiferromagnetic order driven by the superexchange interaction between the local moments \cite{AndersonSuper}. Given the significant separation between the Ce moments, this scenario seems very unlikely.  

\section{Conclusions}
In summary, motivated to get a deeper insight into the factors contributing to the absence of long range order in two Ce-based cage compounds, we performed specific heat measurements over the temperature range $0.39 \le T \le 35$ K and in magnetic fields ranging from 0 to 14 T. By fitting the heat capacity across the whole temperature range and performing the entropy calculations, we found that the ground state multiplet in both materials to be a Kramers doublet with a quartet being the first excited state. We subsequently used our findings to determine the energy level splitting of the Kramers doublet in the ground and its dependence on magnetic field. Employing earlier theoretical calculations, we confirmed that the RKKY interactions between the neighboring local moments must be zero at temperatures smaller than the energy separating the ground state doublet and the first excited state quartet. 

Our findings provide a clear indication as to why no long-range order has been experimentally observed in these materials at temperatures as low as $T=0.138$ K \cite{White2015}, (i) the separation between the Ce ions is too large for antiferromagnetic ordering mediated by the superexchange interactions to appear and (ii) the small hybridization between the $d$- and $f$-orbital states would have produced long-range order.
In this regard, we note that applying pressure to these systems will enhance the hybridization strength and, as a result, superconductivity mediated by valence fluctuations may occur.  Finally, an apparent absence of long range magnetic order in these materials offers an interesting avenue for exploring the interplay between strong coupling superconductivity mediated by the electron-phonon interactions and emerging Kondo lattice coherence. 

\section{Acknowledgments} 
The work at Kent State University was supported by the National Science Foundation grants NSF-DMR-1904315 (A.M. K. and C. C. A.) and NSF-DMR-2002795 (M.D.). The work at UCSD was supported by the US Department of Energy, Office of Basic Energy Sciences, under Grant No. DE-FG02-04ER45105 (single crystal growth) and the National Science Foundation under Grant No. NSF0DMR-1810310 (materials characterization). A.M.K. and Y.Z. contributed equally to this work.

\newpage
\bibliography{Ref_Ce}
\end{document}